\documentclass[aps,prd,twocolumn,preprintnumbers,superscriptaddress,dblfloatfix,nofootinbib]{revtex4-1}
\usepackage[utf8]{inputenc}
\usepackage{lmodern}
\usepackage{amsmath,amssymb}
\usepackage{graphicx}
\usepackage{url}
\usepackage{color}
\usepackage{enumitem}
\usepackage{subfigure}
\usepackage[dvipsnames]{xcolor}
\usepackage[colorlinks=true,breaklinks=true]{hyperref}
\hypersetup{allcolors=[rgb]{0.0 0.0 0.6},linkcolor=[rgb]{0.75 0.05 0.05}}
\usepackage{bm}
\usepackage{epsfig}
\usepackage{amsmath}
\usepackage{amssymb}
\usepackage{slashed}
\usepackage{color}
\usepackage{accents}
\usepackage[dvipsnames]{xcolor}
\usepackage[colorlinks=true,breaklinks=true]{hyperref}
\hypersetup{allcolors=[rgb]{0.0 0.0 0.6},linkcolor=[rgb]{0.75 0.05 0.05}}

\renewcommand\[{\left[}

\newcommand{\TBH}{T_{\rm PBH}}
\newcommand{\MBH}{M_{\rm PBH}}

\newcommand{\rR}{\rho_R}

\newcommand{\exclude}[1]{}

\definecolor{offblue}{RGB}{23,80,153}

\begin{document}

\preprint{IPMU23-0028, IPPP/23/48}

\title{Testing high scale supersymmetry via second order gravitational waves}
	
\author{Marcos M.  Flores} 
\affiliation{Department of Physics and Astronomy, University of California, Los Angeles \\ Los Angeles, California, 90095-1547, USA}
\author{Alexander Kusenko} 
\affiliation{Department of Physics and Astronomy, University of California, Los Angeles \\ Los Angeles, California, 90095-1547, USA}
\affiliation{Kavli Institute for the Physics and Mathematics of the Universe (WPI), UTIAS \\The University of Tokyo, Kashiwa, Chiba 277-8583, Japan}
\affiliation{Theoretical Physics Department, CERN, 1211 Geneva 23, Switzerland}
\author{Lauren Pearce}
\affiliation{Pennsylvania State University-New Kensington, New Kensington, PA 15068}
\author{Yuber F. Perez-Gonzalez}
\affiliation{Institute for Particle Physics Phenomenology, Durham University, South Road, Durham DH1 3LE, United Kingdom}
\author{Graham White}
\affiliation{School of Physics and Astronomy, University of Southampton,\\
	Southampton SO17 1BJ, United Kingdom}

\date{\today}
	
\begin{abstract}
Supersymmetry predicts multiple flat directions, some of which carry a net baryon or lepton number. Condensates in such directions form during inflation and later fragment into Q-balls, which can become the building blocks of primordial black holes.  Thus supersymmetry can create conditions for an intermediate matter-dominated era with black holes dominating the energy density of the universe.  Unlike particle matter, black holes decay suddenly enough to result in an observable gravitational wave signal via the poltergeist mechanism.  We investigate the gravitational waves signatures of supersymmetry realized at energy scales that might not be accessible to present-day colliders. 
\end{abstract}

\maketitle
	

\section{Introduction}

Supersymmetry (SUSY) remains one of the most well motivated paradigms to extend the Standard Model. Although the paradigm conjectures at least twice the number of particles, it is minimal in the sense that it introduces a single extra symmetry for the benefit of solving the hierarchy problem and explaining the asymmetry between matter and anti-matter either through electroweak baryogenesis or the Affleck-Dine mechanism~\cite{Affleck:1984fy, Dine:1995kz, Dine:2003ax}, producing viable dark matter candidates - either the lightest supersymmetric partner \cite{Martin:1997ns} or primordial black holes \cite{Cotner:2016cvr, Cotner:2017tir, Cotner:2019ykd, Flores:2021jas} and naturally predicting a myriad of ``flat directions'' in field space~\cite{Gherghetta:1995dv}. That such a simple framework can provide a unified theory of particle cosmology makes it an attractive paradigm even if hopes are fading that SUSY can solve the hierarchy problem. Moreover, string theory seems to require supersymmetry to be true in some form. The lack of evidence at the LHC motivates considering higher scales of supersymmetry breaking. In this article, we suggest a path to testing such scenarios via the search for a gravitational wave (GW) background.

We will make use of the fact that numerous supersymmetric PBH models predict $M_{\rm PBH}\propto 1/\Lambda_{\rm SUSY}^{2}$, where $\Lambda_{\rm SUSY}$ is the SUSY breaking scale, to explore high-scale SUSY. Light PBHs, corresponding to a high SUSY scale, can cause an early period of matter domination. Generically, black holes evaporate due to Hawking radiation at the event horizon. The black hole evaporation rate is related to the surface to volume ratio and will accelerate once it begins. As the evaporation of black holes is very sudden, the transition from matter domination to radiation will also be rapid. Scalar perturbations that grow during the matter domination do not have time to dissipate and instead produce sound waves that resonantly enhance an otherwise meagre abundance of GWs due to inflation. This mechanism is known as the poltergeist mechanism \cite{Inomata:2019ivs,inomata2019gravitational,Dalianis:2020gup,Domenech:2020ssp,Bhaumik:2020dor,Inomata:2020lmk,Harigaya:2023ecg,Borah:2023iqo,Gehrman:2023esa,Domenech:2023mqk,Kasuya:2022cko,Bhaumik:2022zdd,Chen:2022dah,ReyIdler:2022unr,Bhaumik:2022pil,Kozaczuk:2021wcl,White:2021hwi,Haque:2021dha,Domenech:2021wkk}.

In this work we will explore the possibility of probing high-scale SUSY through the evaporation of light PBHs. In Sec.~\ref{sec:SUSY-QBallFormation} we will review the formation of a scalar condensate and its fragmentation into Q balls. In Sec.~\ref{sec:PBHForm}, we will review the results of Ref. \cite{Flores:2021jas}, discuss the formation of light PBHs from the SUSY Q balls due to long-range scalar forces. In Sec.~\ref{sec:Poltergeist} we discuss details of Hawking radiation, evaporation and evolution of the Universe after the formation of PBHs. Following this, we discuss the poltergeist mechanism, i.e., the GW signal produced from the rapid transition from matter domination to radiation domination which follows shortly after the evaporation of the population of light PBHs.
Throughout the manuscript, we consider natural units where $\hbar = c = k_{\rm B} = 1$, and define the Planck mass to be $M_{\rm Pl}=1/\sqrt{8 \pi G}$, with $G$ being the gravitational constant.

\section{SUSY Background \& Q ball formation}

\label{sec:SUSY-QBallFormation}

The formation of a supersymmetric scalar condensate and its subsequent fragmentation is a well understood phenomenon~\cite{Kusenko:1997si}. The scalar potential $U_{\rm SUSY}$ resulting from a superpotential $W$ can be written as
\begin{equation}
U_{\rm SUSY}
=
\sum_i|F_i|^2
+
\sum_a \frac{g_a^2}{2}|D_a|
\end{equation}
where $F_i = \partial W/\partial \phi_i$, $D_a = \phi_i^\dagger T_a^{ij}\phi_j$ and the sum runs over all the chiral superfields $\phi_i$ and all the gauge generators $T_a$. This potential has flat directions for the linear combinations of fields that make both the $F$-terms and $D$-terms vanish. Inclusion of supersymmetry breaking lifts the flat directions in the potential via the soft symmetry breaking terms, $U_{\rm soft}$, gives
\begin{equation}
U(\phi)
=
U_{\rm SUSY} + U_{\rm soft}
.
\end{equation}
The soft SUSY breaking terms do not include any quartic terms, thus implying that the flat direction remains slowing growing after being lifted. The presence of these flat directions allows for the development of large VEVs during cosmological inflation. This can occur for two reasons: (i) the effective minimum of the potential can be displaced due to terms $\propto H^2\varphi^2$ that come from the Kähler potential~\cite{Affleck:1984fy,Dine:1995kz} and (ii) the field will undergo quantum fluctuations away from the minimum~\cite{Chernikov:1968zm,Tagirov:1972vv,Bunch:1978yq,Linde:1982uu,Lee:1987qc,Starobinsky:1994bd}. Regardless of the cause, after inflation ends and the Hubble parameter decreases, the scalar condensate relaxes. The condensate is subject to instabilities that lead to the formation of Q-balls~\cite{Kusenko:1997si,Dine:2003ax,Enqvist:2003gh}.

The generation and subsequent interactions of the Q-balls is described in~\cite{Flores:2021jas} in detail. As in Ref.~\cite{Flores:2021jas} the physical properties of the Q-balls can be described in terms of the global charge $Q$:
\begin{equation}
M_Q\sim\Lambda |Q|^\alpha,
\quad
R_Q\sim \frac{|Q|^\gamma}{\Lambda},
\quad
\omega_Q\sim\Lambda Q^{\alpha - 1},
\end{equation}
where $M_Q$ and $R_Q$ are the mass and radius of a soliton with charge $Q$ and $\omega_Q$ is the energy per charge. The parameters $\alpha$ and $\gamma$ are positive and less than or equal to one, and $\Lambda$ is the energy scale associated with the scalar potential. For a flat direction lifted by  gauge-mediated supersymmetry breaking, $\alpha = 3/4$ and $\gamma = 1/4$~\cite{Dvali:1997qv,Kusenko:1997si}. Different supersymmetry breaking mechanisms can lead to different values of $\alpha$ and $\gamma$~\cite{Enqvist:2003gh}. 

Crucial to our mechanism will be the interaction amongst Q-balls. For simplicity, we will focus on a simple model that encapsulates the physics required to describe Q-ball interactions. We assume interactions are mediated by a light field $\chi$, which can be a second flat direction in field space.  In particular, we assume that these two flat directions $\phi$ and $\chi$  are lifted by some soft terms from a gauge-mediated supersymmetry breaking mechanism~\cite{deGouvea:1997afu,Dvali:1997qv,Hong:2017ooe,Kawasaki:2019ywz,Kasuya:2000sc}:

\begin{align}
U_{\rm soft}(\phi) &\approx \Lambda^4
\left[
\log\left(
1 + \frac{|\phi|^2}{M^2}
\right)
\right]^2 + U_{\rm grav} + \cdots\\[0.25cm]
U_{\rm soft}(\chi) &\approx \Lambda^4
\left[
\log\left(
1 + \frac{|\chi|^2}{M^2}
\right)
\right]^2 + U_{\rm grav} + \cdots
.
\end{align}
Following Ref.~\cite{Flores:2021jas}, we will assume that gauge mediated terms dominate over the gravity mediated terms, so that $U_{\rm grav}$ can be neglected. Here, $\Lambda$ is the scale associated with supersymmetry breaking, and in the unbroken limit $\Lambda\to 0$. In our discussions below, $\Lambda$ will correspond to the height of the potential $U_{\rm soft}$.

The two flat directions can be coupled through higher dimensional operators. This leads to interaction terms of the form
\begin{equation}
V_{\chi\phi}(\phi,\chi) = -y\chi\phi^\dagger\phi + {\rm h.c.},
\end{equation}
where $y$ is an effective, dimensionful coupling. It is this interaction which generates an attractive force between Q-balls, and ultimately, results in PBHs~\cite{Flores:2021jas}. 

We also note that the field $\chi$ can receive an effective mass $m_\chi\sim H$, of the order of the Hubble parameter~\cite{Kawasaki:2011zi}, on length scales smaller than $m_\chi^{-1}\sim H^{-1}$, and so it generates a long-range attractive force.

\section{PBH formation}

\label{sec:PBHForm}

After the formation of Q-balls, the charges act similarly as a system of fermions interacting via Yukawa interactions~\cite{Amendola:2017xhl,Savastano:2019zpr,Flores:2020drq,Domenech:2021uyx,Flores:2021tmc, Flores:2023zpf}. The system can form bound halos during radiation or matter dominated eras and will be subject to radiative cooling by emission of $\chi$ waves. Energy can be removed a variety of waves, although pairwise interactions of charges is expected to be the main contributor to removing energy from a given system of charges (that is, bremsstrahlung emission dominates). The characterstic time scale associated with collapse of a system of charges is given by
\begin{equation}
t_{\rm cool}
=
\frac{E}{dE/dt}
=
\frac{E}{P_{\rm brem} + \cdots}
\end{equation}
where the dots indicate additional radiative channels, including coherent or incoherent oscillatory sources of radiation. When $t_{\rm cool} < H^{-1}$ radiative cooling is efficient and a system of Q-balls will begin to collapse. As extended objects, the charges will begin to merge once the Q-ball halo reaches a radius $R = N^{1/3}R_Q$. This processes is rapid and will lead to a single charge per horizon with charge $N\langle Q\rangle$.

As the horizon increases these newly formed larger Q-balls will be able to interact, initiating another period of mergers. This processes will continue, generating larger and larger Q-balls. If the scalar potential responsible for Q-ball formation is flat indefinitely, the VEV inside a given Q-ball can exceed the Planck scale. To avoid reaching a scale which is inconsistent with our field-theoretical description of Q-balls, we note that higher dimensional operators may play an important role as the VEV increases. In particular operators of the form~\cite{Kusenko:2005du} 
\begin{equation}
V^n(\phi)_{\rm lifting}
\approx
\lambda_n M^4	
\left(
\frac{\phi}{M}
\right)^{n - 1 + m}
\left(
\frac{\phi^*}{M}
\right)^{n - 1 - m},
\end{equation}
lift the flat direction potential, altering the evolution and growth of the Q-balls. Here $M$ is the scale of new physics, i.e., $M_{\rm GUT}$ or $M_{\rm Pl}$ and $\lambda_n\sim \mathcal{O}(1)$. When the VEV reaches some critical value, $Q_c$, these operators transition the Q-balls from flat-direction Q-balls ($\alpha = 1/4$, $\gamma = 3/4$) into Coleman Q-balls ($\alpha = 1$, $\gamma = 1/3$) which have $Q$-independent VEVs. Explicitly,
\begin{equation}
Q_c
\simeq
\lambda_n^{-\frac{2}{n - 1}}
\left(
\frac{M}{\Lambda}
\right)^{\frac{4n - 12}{n - 1}}.
\end{equation}
For $\Lambda\sim 10^3$ GeV and $M\sim 10^3$ GeV, $Q_c\sim 10^{17}$ for $n = 4$ or as large as $10^{32}$ for $n = 7$. The Q-ball evolution and growth is also affected by the decay of the Q-ball charge~\cite{Kawasaki:2005xc}. We note that, for $m\neq 0$, the above operators can also cause decay of the global charge inside the Q-ball. 

After reaching the critical charge, mergers continue until the maximum allowed charge is reached. This maximum is determined by the equality $R_{s} = 2GM_Q = R_Q$ which implies,
\begin{equation}
Q_{\max}
=
\frac{2^{3/4}\sqrt{3}}{\sqrt{\pi}}Q_c^{1/4}
\left(
\frac{M_{\rm Pl}}{\Lambda}
\right)^3
.
\end{equation}
Once mergers result in charges equal to $Q_{\max}$, the Q-ball collapses into itself and a PBH is formed. The newly formed black holes no longer participate in scalar interactions in accordance with the no-hair theorems. 
The mass of the resulting PBH is given by~\cite{Kusenko:2005du, Flores:2021jas}
\begin{equation}
\begin{split}
\MBH
&=
\omega_c Q_{\max}
=
\left(
\pi\sqrt{2}\Lambda Q_c^{-1/4}
\right)
Q_{\max}\\
&\sim 
2\times 10^{4}\ {\rm g}
\left(
\frac{10^{14}\ {\rm GeV}}{\Lambda}
\right)^2
\end{split}
\end{equation}
PBHs of this mass occur when the global charge within a critical radius, $R_*$, is collected in one Q-ball with size $Q_{\max}$,
\begin{equation}
Q_{\max} = \frac{4\pi}{3}q_0R_*^{3}
\end{equation}
where $q_0$ is is the charge density at fragmentation. The ratio,
\begin{equation}
\frac{R_*}{H_f^{-1}}
=
\left(
\frac{Q_{\max}}{N Q_0}
\right)^{1/3}
\end{equation}
characterizes how much larger the Hubble radius is, relative to the Hubble radius at fragmentation, so that 
$\MBH = \omega_cQ_{\max}$. 
Here, we assumed that $N$ charges of average charge $Q_0$ formed within the horizon at $H_f$. The product $N Q_0$ can be related to the charge asymmetry left in the condensate at the time of fragmentation through
\begin{equation}
N Q_0
=
\frac{4\pi}{3}
H_f^{-3} \eta_Q T_f^3
.
\end{equation}
The flat direction we specified may or may not be the one associated with the Affleck-Dine baryogenesis. If they are unrelated, then $\eta_Q$ may take any value besides $\eta_B\sim 10^{-10}$. For later use, we note that this implies that
\begin{equation}
\label{eq:RstarEtaQ}
R_*
=
\frac{1}{T_f}\left(
\frac{3Q_{\max}}{4\pi \eta_Q}
\right)^{1/3}
.
\end{equation}

We may consider the case when the same flat direction is used for generating both the baryon asymmetry and a population of light PBHs. Initially the condensate starts at with $\eta_{Q,i} \geq \eta_B$ at some high temperature. The condensate will decay into fermions at the rate $\Gamma_Q$ such that at the time of fragmentation, $\eta_Q = \eta_{Q,i}\exp(-\Gamma_Q H_{f}^{-1})$. Any fermion coupled to the flat direction will be very massive, implying that the decay rate is exponentially suppressed, $\Gamma_Q\propto \exp(-1/\varepsilon)$, where $\varepsilon = \omega/(g\left\langle\phi\right\rangle)\ll 1$ where $\omega$ is the energy density per unit charge in the condensate~\cite{Dolgov:1989us,Pawl:2004cs}.

The abundance of PBHs at formation is given by
\begin{equation}
\rho_{\rm PBH}(a_*)
=
\frac{3\MBH}{4\pi H_*^{-3}},
\end{equation}
where $a_*$ is the scale factor at formation. We will compare this to the critical density and define $\beta = \rho_{\rm PBH}(a_*)/\rho_{\rm crit}$. Using \eqref{eq:RstarEtaQ},
\begin{equation}\label{eq:beta}
\beta
=
\frac{T_f}{\Lambda}
\left(
\frac{\eta_Q}{2Q_{c}^{1/4}}
\right)^{1/3}
.
\end{equation}
Rather than specifying the asymmetry $\eta_Q$, we can specify the initial abundance $\beta$ and instead deduce the required asymmetry required to obtain such a fraction. In particular, we make use of the fact that
\begin{equation}
\Gamma_Q 
\sim 
e^{-1/\varepsilon}
\frac{\Lambda^4}{\eta_Q T_f^3},
\qquad
\frac{\Gamma_Q}{H_f} = \ln(\eta_B/\eta_Q).
\end{equation}
It follows that the asymmetry $\eta_Q$ is given by
\begin{equation}
\eta_Q
=
\eta_B
\exp
\left[
-
\frac{\Lambda e^{-1/\varepsilon}}{2H_f\beta^3 Q_c^{1/4}}
\right]
.
\end{equation}
There are a number of constraints which will limit the accessible parameter space. First, we will require that the energy density in the initial population of charges is less than the critical density,
\begin{equation}
N < \frac{4\pi M_{\rm Pl}^2}{\Lambda Q_0^\alpha H_f}
.
\end{equation}
%
\begin{figure*}[!t]
\centering
    \includegraphics[width=0.8\linewidth]{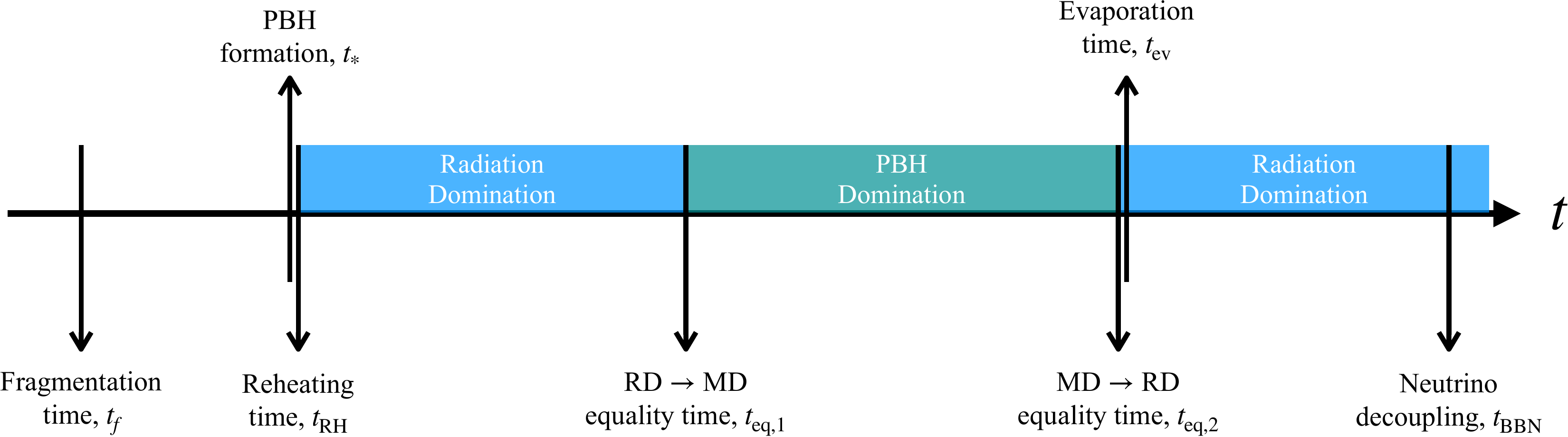}
    \caption{\label{fig:times} Overview of timeline in our scenario.}
\end{figure*}
Second, we require that the potential energy density due to Yukawa interactions is less than the critical density,
\begin{equation}\label{eq:Yuklim}
N < \sqrt{4\pi}
\left(
\frac{\Lambda M_{\rm Pl}}{y H_f Q_0^{2-\alpha}}
\right)
.
\end{equation}
Third, we require that all of the resulting Q-balls formed by the charges can fit within the horizon at fragmentation
\begin{equation}
N < \frac{\Lambda^3}{Q_0^{3\gamma}H_f^3}
.
\end{equation}
In what follows, we fix $N=10^6$.
Fourth, we demand that the fragmentation temperature is greater than height of the flat direction, i.e. $1 < T_f/\Lambda$. Lastly, as we mentioned previously, this processes will occur before the end of reheating. To ensure that the long-range scalar force is not limited by thermal screening, we require that PBHs form before reheating ends. This leads to the condition
\begin{equation}
T_{\rm RH}
\lesssim
2^{5/24}
\left(\frac{\pi}{3}\right)^{1/4}
\left[
\frac{e^{1/\varepsilon}\Lambda^7}{H_f Q_c^{1/4}\ln(\eta_B/\eta_Q)}
\right]^{1/6}
.
\end{equation}

\bigskip

\section{Poltergeist Formalism}

\label{sec:Poltergeist}


Once formed, and depending only on its initial characteristics, the PBH starts emit particles, including those beyond the Standard Model, via Hawking radiation.  The PBHs formed from Q-ball merging are Schwarzschild black holes, with a horizon mass and temperature determined by the instantaneous mass $\MBH$,
\begin{align}
	 r_S = \frac{\MBH}{4\pi M_{\rm Pl}^2}, \quad \TBH = \frac{M_{\rm Pl}^2}{\MBH}.
\end{align}
As a result of the particle emission, black holes lose mass over time at a rate that depends on the spectrum of degrees of freedom existing in nature,
\begin{align}\label{eq:Mrate}
	 \frac{d\MBH}{dt}  &= -\sum_{i=\text{all particles}}\, \frac{g_i}{2\pi}\int_{m_i}^\infty  \frac{\vartheta_{s_i}(\MBH, E) E dE}{\exp(E/\TBH) - (-1)^{2s_i}} \notag\\
	 &= - \varepsilon(\MBH)\, \frac{M_{\rm Pl}^4}{\MBH^2}
\end{align}
where $g_{i}$,  $m_i$, and $s_{i}$ denote the number of internal degrees of freedom (dof), mass, and spin of particle $i$, respectively. 
The $s_i$-dependent factors, known as greybody factors, $\vartheta_{s_i}(\MBH,E)$ describes possible back-scattering due to the centrifugal and gravitational forces in the curved spacetime around the black hole.
Finally, in Eq.~\eqref{eq:Mrate}, we defined the \emph{evaporation} function $\varepsilon(\MBH)$, which contains the information of the particle spectrum that can be emitted including all SUSY degrees of freedom, see Ref.~\cite{Cheek:2021odj} for further details.
For the specific scenario that we are considering, we assume a minimal spectrum consisting of all SM degrees of freedom and assume that all SUSY degrees-of-freedom have masses equal to the scale $\Lambda$. 

The lifetime $\tau$ of any PBH will then depend on its initial mass and the particle spectrum, and can be determined by integrating the lost mass rate in Eq.~\eqref{eq:Mrate}.
A common approximation consists in considering the denominated \emph{geometrical optics} for the greybody factors, $\vartheta_{s_i}(\MBH, E) \to 27 \MBH^2 E^2/16 \pi^2 M_{\rm Pl}^4$.
Using such an approximation for the mass depletion rate, we can estimate the lifetime $\tau$ as
\begin{align}
    \tau = \frac{640}{27 \pi g_{\rm \star, BH}}\frac{M_h^3}{M_{\rm Pl}^4},
\end{align}
where $g_{\rm \star, BH} = 106.75$ the number of degrees of freedom of the SM~\cite{Baldes:2020nuv}.
We stress, however, that the results presented in the next subsections are obtained including the full greybody factors, and determining carefully the PBH evaporation.

Before completely vanishing, the PBH fraction of total Universe energy density increases proportional to the scale factor $a$ since PBHs behave as matter, $\rho_{\rm PBH} \propto a^{-3}$.
Thus, an early matter dominated era could have occurred provided that the PBHs do not evaporate prior to a matter-radiation equality.
A simple criterion to establish whether a PBH dominated era could have taken place is obtained by comparing the plasma temperature at which the evaporation would have occurred $T_{\rm ev}$ with plasma temperature at formation, assumed to be the reheating temperature $T_{\rm RH}$~\cite{Hooper:2019gtx}
\begin{align}
	\beta \gtrsim \beta_{\rm dom} \equiv \frac{T_{\rm ev}}{T_{\rm RH}}.
\end{align}
To consistently track the evolution of the Universe after the PBH formation, we solve numerically the set of Friedmann and Boltzmann equations for the comoving PBHs ($\varrho_{\rm BH} \equiv a^3 \rho_{\rm BH}$) and radiation ($\varrho_{\rm R} \equiv a^4 \rR$) and energy densities as function of the scale factor~\cite{Gutierrez:2017ibk,Cheek:2021odj},
\begin{subequations}\label{eq:UnEv}
\begin{align}
	H^2 &=\frac{1}{3\, M_P^2} \left(\varrho_{\rm BH} a^{-3}+\varrho_{\rm R} a^{-4}\right)\,,\\
 	\frac{d\varrho_{\rm R}}{d \ln a} &= -\frac{{\rm BR_{BH\to SM}}}{H}\frac{d\ln\MBH}{dt}a\varrho_{\rm BH}\,,\label{eq:UnEvRad}\\ 
 	\frac{d\varrho_{\rm BH}}{d \ln a} &=\frac{1}{H}\frac{d\ln\MBH}{dt} \varrho_{\rm BH}\,,
\end{align}
\end{subequations}
where ${\rm BR_{BH\to SM}} \equiv \varepsilon_{\rm SM}(\MBH)/\varepsilon(\MBH)$ denotes the ``Branching Ratio'' of the PBH evaporation into SM particles.
The evolution of the energy densities together with the evolution of the PBH mass, Eq.~\eqref{eq:Mrate}, is traced with the help of the python package {\tt FRISBHEE}~\cite{Cheek:2022dbx,Cheek:2022mmy}, modified to account for the specific PBH formation mechanism considered here.
\begin{figure*}[!t]
\centering
    \includegraphics[width=\linewidth]{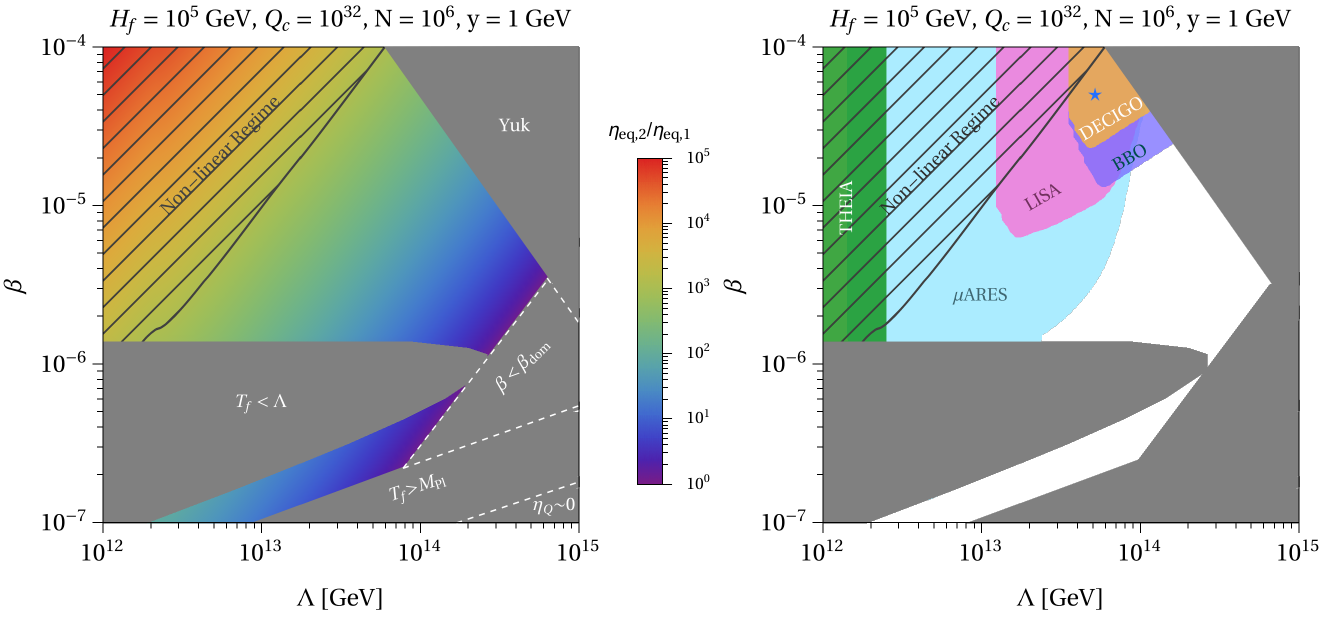}
    \caption{\label{fig:senst} (Left) Ratio of the conformal time at evaporation $\eta_{\rm eq, 2}$ to that at PBH-radiation equality $\eta_{\rm eq, 1}$ for the parameter space spanned by the initial PBH fraction $\beta$ and the scale $\Lambda$. The shaded regions are constrained by different requirements (see text), and the hatched region corresponds to the non-linear regime for the GW generation in the poltergeist mechanism. (Right) Sensitivity to the same parameter space from future GW observatories, THEIA (green), $\mu$-ARES (light blue), LISA (pink), DECIGO (darker orange), BBO (purple). The blue star indicates the benchmark chosen for Fig.~\ref{fig:GWs}.}
\end{figure*}

Curvature perturbations source gravitational waves at second order in perturbation theory~\cite{Assadullahi:2009nf,Alabidi:2013lya,Kohri:2018awv}.  The strength of the resulting gravitational wave signal depends on the cosmological history of the universe through the gravitational potential.  During an early matter dominated epoch, the gravitational potential does not decay, even at subhorizon scales.  Therefore, the resulting gravitational wave spectrum is enhanced as compared to that of a universe without an early matter dominated epoch.  Furthermore, if the matter dominated period ends abruptly, then there is a resonant-like enhancement~\cite{Inomata:2019ivs,Inomata:2020lmk}.  This can be understood as a consequence of the PBH overdensities converting into relativistic sound waves when the PBHs decay, and gravitational waves comoving with these sound waves under go rapid enhancement~\cite{Ananda:2006af,Inomata:2019ivs,Inomata:2020lmk}.  Note that if the transition between matter and radiation domination is slow, the overdensities gradually dissolve instead of producing relativistic sound waves.

This poltergeist phenomenon has been studied particularly in the context of PBH evaporation in~\cite{Inomata:2020lmk}.  Because we consider a population of black holes with a narrow mass distribution, our analysis follows the instantaneous transition limit of~\cite{Inomata:2019ivs}.  The time-averaged gravitational wave power spectrum is
\begin{align}
\overline{{\cal P } _h (\eta ,k)} = 4 \int _0 ^\infty dv & \int _{|1-v|}^{1+v} du \left( \frac{4 v^2-(1+v^2 - u ^2)^2}{4 v u} \right) ^2 \nonumber \\ 
& \times \overline{I^2(u,v,k,\eta,\eta _R} {\cal P} _\zeta (uk) {\cal P} _\zeta (vk) \ ,
\end{align}
where $P_\zeta = A_s ( k \slash k_*)^{n_s-1}$ is assumed to result from an earlier inflationary epoch.  The gravitational wave signal at some conformal time $\eta$ is related to the power spectrum by 
\begin{equation}
    \Omega _{\rm GW} (\eta , k) = \frac{1}{24} \left( \frac{k}{a (\eta )H(\eta )} \right)^2 \overline{{\cal P } _h (\eta ,k)}.
\end{equation}

The impact of the early matter dominated epoch appears in the time evolution of 
\begin{equation}
    I(u,v,k,\eta,\eta _R) = \int _0 ^{k \eta} d (\overline{k \eta}) \frac{a (\bar{\eta})}{a(\eta)} k G_k (\eta , \bar{\eta} ) f(u,v, \overline{k \eta} , k \eta _R),
\end{equation}
which involves the source function
\begin{align}
f(u,v, \overline{k \eta} , k \eta _R) &= \frac{3}{25(1+w)} \left( 2(5+3 w) \Phi (u \overline{k \eta} ) \Phi (v \overline{ k \eta})  \right.  \nonumber \\ 
& \left. + 4 H^{-1} \frac{\partial}{\partial \eta } \left(  \Phi (u \overline{k \eta} ) \Phi (v \overline{ k \eta}) \right) \right. \nonumber \\ & \left. +4 H^{-2} \frac{\partial}{\partial \eta }   \Phi (u \bar{k \eta} )\frac{\partial}{\partial \eta }   \Phi (v \overline{k \eta} ) \right) \ .
\end{align}
as well as the Green's function which solves 
\begin{equation}
    \frac{\partial ^2 G(\eta , \bar{\eta})}{\partial \eta ^2} + \left( k^2 - \frac{1}{a}\frac{\partial ^2 a}{\partial \eta ^2} \right)G(\eta , \bar{\eta})  = \delta (\eta - \bar{\eta } ) .
\end{equation}

The Green's function is different during matter and radiation domination as the scale factor $a$ and equation of state $w$ are different, but the key difference is in the source function, as the gravitational potential solves 
\begin{equation}
    \frac{\partial ^2 \Phi}{\partial \eta ^2} + 3(1+w) H \frac{\partial \Phi }{\partial \eta } + w k^2 \Phi = 0 \ .
\end{equation}
During radiation domination this decays away at subhorizon scales whereas it is constant during matter domination.

For a sufficiently rapid transition from matter to radiation domination, we can use the approximate analytic formulas for the gravitational wave power spectrum in Ref.~\cite{Inomata:2019ivs}. 

If the end of the matter dominated epoch is sufficiently late, matter perturbations may enter the non-linear regime.  In this case, we follow the conservative approach of Ref.~\cite{Inomata:2019ivs}, and consider modes only up to $k_{\rm max} = 470 \slash \eta_R$, where $\eta_R$ is the conformal time at the end of the matter dominated epoch.

\section{Results \& Discussion}

After determining the full PBH evolution and the subsequent generation of GWs as described above, we estimate the sensitivity of future GW detectors by computing the signal-to-noise ratio, defined by~\cite{Smith:2019wny}
\begin{align}
    {\rm SNR}^2 = T \int_{f_{\rm min}}^{f_{\rm max}} df \left(\frac{\left.\Omega h^2\right|_{\rm polter}}{\left.\Omega h^2\right|_{\rm sens}}\right)^2,
\end{align}
where $\left.\Omega h^2\right|_{\rm polter}$ is the GW spectrum produced by the PBH sudden evaporation, $\left.\Omega h^2\right|_{\rm sens}$ is the experimental sensitivity, and $T$ its expected exposure.
We consider the future proposals of THEIA~\cite{Theia:2017xtk,Garcia-Bellido:2021zgu}, with $T=20$ years, $\mu$ARES~\cite{Sesana:2019vho}, having $T=7$ years, LISA~\cite{Caprini:2019egz} with $T=4$ years, 3-units DECIGO~\cite{Moore:2014lga,Kawamura:2020pcg} with $T=1$ year, and BBO~\cite{Cutler:2009qv} having a $T=5$ year observation period.
We require the signal-to-noise ratio to exceed ${\rm SNR}=5$ for an experiment to be considered sensitive to the particular parameters associated.

In Fig.~\ref{fig:senst}, we depict the ratio of the conformal time at evaporation $\eta_{\rm eq, 2}$ to that at PBH-radiation equality $\eta_{\rm eq, 2}$ (left panel), and the sensitivity of the aforementioned GW experiments (right panel) to the parameter space defined by the initial PBH fraction $\beta$ and the SUSY breaking scale $\Lambda$.  We note that the poltergeist mechanism is extremely sensitive to the ratio of $\eta_{\rm eq , 2} \slash \eta_{\rm eq, 1}$; the analytic fits in Ref.~\cite{Inomata:2019ivs} has the seventh power of this ratio.  Therefore the signal is strongest in the upper left corner, which is thus the regime best probed by future experiments.  
For these figures, we assumed the values of $H_f = 10^5$ GeV for the Hubble radius at the time of Q-ball fragmentation, $Q_c =10^{32}$, and $y=1$ GeV.
In both panels, the gray regions correspond to parameters restricted by previously mentioned constraints, specifically: $T_f<\Lambda$, $T_f>M_{\rm Pl}$, regions where the fragmentation temperature is either smaller than the height of the flat direction or larger than the Planck scale, respectively, $\eta_Q\sim 0$ where the charge asymmetry becomes suppressed, and the area marked as ``Yuk" where the potential energy density surpasses the critical density (refer to eq.~\eqref{eq:Yuklim}).
Additionally, the region marked with $\beta < \beta_{\rm dom}$ indicates the parameters where PBHs are formed, but they do not dominate the evolution of the Universe prior evaporation.
The hatched region in both panels denotes the non-linear regime in the poltergeist mechanism, characterized by $\eta_{\rm eq, 2}/\eta_{\rm eq, 1} > 470$.

Upon examining the sensitivity of different experiments to the parameters under consideration, we find that observatories capable of measuring the GW background in the high-frequency region $f \gtrsim 10^{-4}$ Hz, i.e., LISA, DECIGO, and BBO, will be able to probe the highest SUSY scales of $\Lambda \gtrsim 10^{13}$ GeV, for initial PBH fractions of $\beta\gtrsim 6 \times 10^{-6}$ for LISA, $\beta \gtrsim 1.3\times 10^{-5}$ for BBO, and $\beta\gtrsim 2\times 10^{-5}$ for DECIGO.
$\mu$-ARES is anticipated to offer a broader range of sensitivity, potentially testing PBH initial fractions as low as $\sim 10^{-6}$ and scales of $\Lambda\lesssim 10^{14}$ GeV.
However, we highlight that a substantial portion of the expected sensitivity region resides within the non-linear regime. Consequently, it is reasonable to expect modifications when considering the complete determination of the GW spectrum. This observation holds particular relevance for THEIA, whose entire sensitivity region falls within the non-linear regime.
\begin{figure*}[!ht]
\centering
    \includegraphics[width=0.95\linewidth]{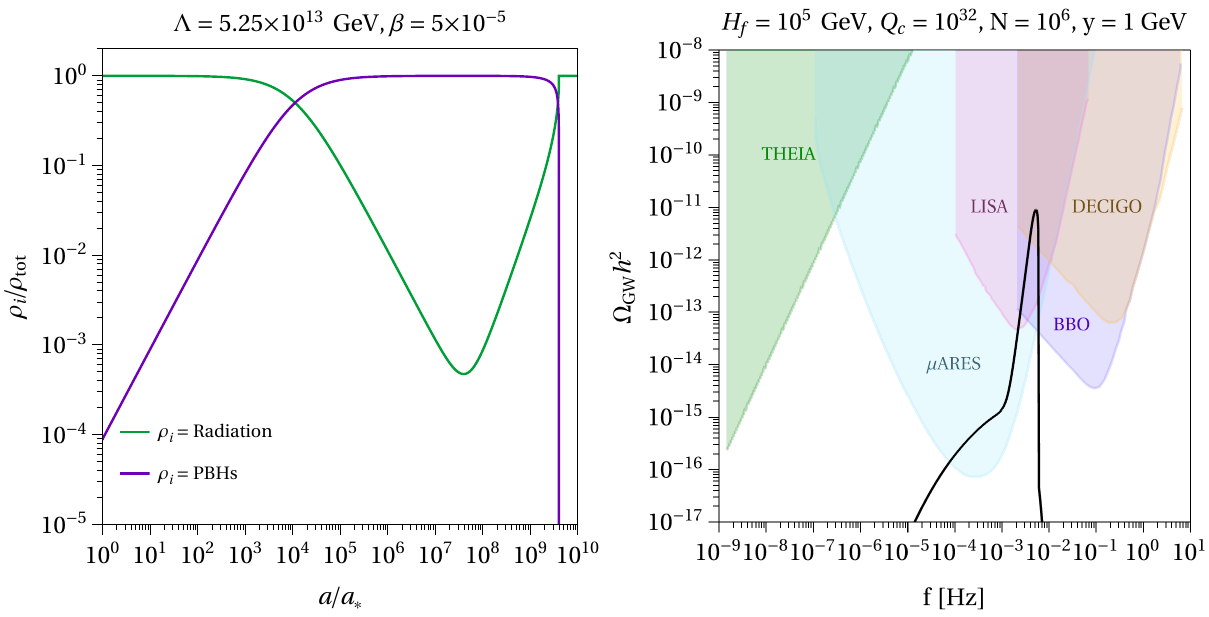}
    \caption{\label{fig:GWs} (Left) PBH (purple) and radiation (green) energy densities normalized to the total as function of the scale factor for benchmark parameters of $\Lambda = 5.25\times 10^{13}~{\rm GeV}$ for the SUSY breaking scale, $\beta = 5\times 10^{-5}$ for the initial PBH fraction. (Right) GW spectrum for the same benchmark (full black) as function of the frequency, together with sensitivity regions for THEIA (green), $\mu$-ARES (light blue), LISA (light red), BBO (light purple), and DECIGO (light orange).}
\end{figure*}

In the right panel of Fig.~\ref{fig:GWs}, we display the GW spectrum corresponding to selected benchmark parameters in Fig.~\ref{fig:senst}, $\Lambda = 5.25\times 10^{13}~{\rm GeV}$ for the SUSY breaking scale, $\beta = 5\times 10^{-5}$ for the initial PBH fraction. 
This scenario is depicted by a solid black line.
We also map the anticipated sensitivity ranges of various upcoming GW detectors, denoted by their specific colors: THEIA (green), $\mu$-ARES (light blue), LISA (light red), BBO (light purple), and DECIGO (light orange).
In the left panel of the same figure, we plot the evolution of the PBH and radiation energy densities, both normalized to the total energy density subsequent to PBH formation. This is depicted as a function of the scale factor, which is normalized with respect to the scale factor at the time of PBH formation.
We observe that the PBH population generated by Q-ball collapse indeed dominates the evolution of the energy density of the Universe at around $a/a_* \sim 10^4$, and evaporated completely at $a/a_* \sim 3 \times 10^9$.
For the chosen benchmark, it is worth noting that the predicted GW spectrum lies with the sensitivity ranges of four out of the five experiments under consideration, $\mu$-ARES, LISA, BBO, and DECIGO. These experiments are anticipated to detect the GW background with a notable degree of sensitivity. Specifically, $\mu$-ARES is expected to provide extensive coverage of this signal, enabling a thorough examination of the spectral shape of the GW background.

Recently, the NANOGrav collaboration provided evidence pointing towards the existence of a stochastic GW signal, with a confidence level (C.L.) of $\sim 4\sigma$~\cite{NANOGrav:2023ctt,NANOGrav:2023gor,NANOGrav:2023hfp,NANOGrav:2023hvm,NANOGrav:2023icp}. 
Given this development, it is natural to inquire if the poltergeist GWs in our proposed scenario could account for such a signal.
Upon analyzing Fig.~\ref{fig:senst} (right panel), it appears that the NANOGrav signal could potentially be attributed to PBHs, formed due to the collapse of Q-balls, especially when the SUSY scale is around $10^{12}$GeV, given that THEIA is expected to cover the same spectrum region as NANOGrav.
However, as mentioned before, THEIA's sensitivity predominantly lies within the non-linear regime. 
Consequently, an accurate prediction for the NANOGrav signal would necessitate a comprehensive understanding of the poltergeist dynamics in this regime.
Given the intricacies involved, we exercise caution and choose not to further pursue this line of interpretation.

\begin{acknowledgments}
The work of YFPG has been funded by the UK Science and Technology Facilities Council (STFC) under grant ST/T001011/1. 
This project has received funding/support from the European Union’s Horizon 2020 research and innovation programme under the Marie Sk\l{}odowska-Curie grant agreement No 860881-HIDDeN.
This work has made use of the Hamilton HPC Service of Durham University. 
A.K. and M.M.F were supported by the U.S. Department of Energy (DOE) Grant No. DE-SC0009937. M.M.F was also supported by the University of California, Office of the President Dissertation Year Fellowship and donors to the UCLA Department of Physics \& Astronomy.  A.K. was also supported by World Premier International Research Center Initiative (WPI), MEXT, Japan, and by Japan Society for the Promotion of Science (JSPS) KAKENHI Grant No. JP20H05853.

\end{acknowledgments}


\bibliography{biblio}


\end{document}